%% file: ms.tex
\documentclass[sigconf]{acmart}

\newcommand{\systemname}{HierCat}

\usepackage[ruled,linesnumbered]{algorithm2e}
\usepackage{multirow}

\AtBeginDocument{%
  \providecommand\BibTeX{{%
    \normalfont B\kern-0.5em{\scshape i\kern-0.25em b}\kern-0.8em\TeX}}}

\copyrightyear{2023}
\acmYear{2023}
\setcopyright{acmlicensed}\acmConference[WWW '23 Companion]{Companion Proceedings of the ACM Web Conference 2023}{April 30-May 4, 2023}{Austin, TX, USA}
\acmBooktitle{Companion Proceedings of the ACM Web Conference 2023 (WWW '23 Companion), April 30-May 4, 2023, Austin, TX, USA}
\acmPrice{15.00}
\acmDOI{10.1145/3543873.3584622}
\acmISBN{978-1-4503-9419-2/23/04}

\begin{document}


\title{{\systemname}: Hierarchical Query Categorization from Weakly Supervised Data at Facebook Marketplace}



\newcommand{\authorStr}{Yunzhong He, Cong Zhang, Ruoyan Kong, Chaitanya Kulkarni, Qing Liu, Ashish Gandhe, Amit Nithianandan and Arul Prakash}

\author{Yunzhong He$^1$, Cong Zhang$^1$, Ruoyan Kong$^2$, Chaitanya Kulkarni$^1$, Qing Liu$^1$, Ashish Gandhe$^1$, Amit Nithianandan$^1$, Arul Prakash$^1$}
\affiliation{%
  \institution{$^1$Meta, $^2$University of Minnesota}
  \country{United States}
}
\email{{yunzhong, conzhang, chaitanya2, qingl, ashigan, anithian, arulprakash}@meta.com}
\email{kong0135@umn.edu}

\renewcommand{\shortauthors}{Yunzhong He}

\begin{abstract}
\input{abstract}
\end{abstract}



\maketitle

\section{Introduction}\label{sec:intro}
\input{intro.tex}

\section{Methodology}\label{sec:model}
\input{methology}

\section{Offline Experiments}\label{sec:offline_experiments}
\input{experiments.tex}

\section{Online Experiments}\label{sec:online_testing}
\input{online_testing.tex}

\section{Related Work}\label{sec:related_work}
\input{related_work.tex}

\section{Conclusion}\label{sec:conclusion}
\input{conclusion.tex}


\bibliographystyle{ACM-Reference-Format}
\bibliography{bibliography}

\end{document}

%% file: abstract.tex
Query categorization at customer-to-customer e-commerce platforms like Facebook Marketplace is challenging due to the vagueness of search intent, noise in real-world data, and imbalanced training data across languages. Its deployment also needs to consider challenges in scalability and downstream integration in order to translate modeling advances into better search result relevance. In this paper we present {\systemname}, the query categorization system at Facebook Marketplace. {\systemname} addresses these challenges by leveraging multi-task pre-training of dual-encoder architectures with a hierarchical inference step to effectively learn from weakly supervised training data mined from searcher engagement. We show that {\systemname} not only outperforms popular methods in offline experiments, but also leads to 1.4\% improvement in NDCG and 4.3\% increase in searcher engagement at Facebook Marketplace Search in two weeks of online A/B testing.

%% file: intro.tex
\begin{table}[tb]
\vspace{2mm}
\begin{tabular}{c|c}
    \toprule
    {\small Query} & {\small Category path}\\
    \midrule
    {\small iPhone cases} & {\small Electronics//Cell Phones//Accessories//Cases}\\
    {\small 2019 brz black} & {\small Vehicles//Cars \& Trucks//Coupes}\\ 
    {\small 1b1b for rent} & {\small  Housing//Property Rentals}\\
    \bottomrule
\end{tabular}
\vspace{2mm}
\caption{Popular queries on Marketplace and their categories}
\label{tab:query_category_examples}
\vspace{-2.0em}
\end{table}

Query categorization refers to the task of mapping a search query to a predefined product taxonomy, which usually consists of a large label space organized in a hierarchical structure. It helps e-commerce search engines to understand users' shopping intents in a structured way to provide high quality search experiences. On customer-to-customer (C2C) shopping platforms like Facebook Marketplace\footnote{http://www.facebook.com/marketplace}, query categorization can be particularly challenging due to the following observations:

\textit{Noisy training data:} Training query categorization models based on <query, product> engagement is a common approach to avoid needing massive amount of labeled data \cite{aprfnet, Liu2019SystemDO, backoffOptimize, dynmicQueryIntentJD}. However, due to how product categories are labeled and the browsy behavior of user, training data mined from search engagement can be heterogeneous and noisy.

\textit{Vague search intent:} Unlike a product listing, whose category is usually well-defined and precise, a search query's category can be vague. For example, query "used electronics" refers to a high-level category "Electronics" which is rarely predicated alone for a product listing.

\textit{Internationalization:} Facebook Marketplace is a global platform and aims to offer consistent search experiences across languages. However, training data mined from search engagement are naturally skewed towards popular languages like English.

In this paper, we present {\systemname}, a query categorization system designed to learn from weakly supervised training data mined from search engagement. Unlike popular query categorization methods based on text classification approach \cite{aprfnet, dynmicQueryIntentJD, Liu2019SystemDO, Skinner2019EcommerceQC, contextAwareQC}, we introduce a dual-encoder architecture with a hierarchical inference step to leverage both the textual information of category labels and the hierarchical structure of product taxonomy. The encoder also incorporates state-of-the-art transformer architecture pre-trained on cross-language data as well as a downstream product retrieval task. We show that this simple and effective approach can substantially reduce the noise in training data, and generalizes to unseen queries better than many popular approaches. We also share our inference-time optimizations like category embedding caching and beam search to speed up inference, and our deployment story around translating modeling advances to better search experiences. {\systemname} is now powering hundreds of millions of search queries on Facebook Marketplace per day, helping users find relevant product listings of their interests.

%% file: methology.tex
\begin{figure*} [!h]
\centering
\vspace{-0.1em}
\includegraphics[width=0.8\textwidth]{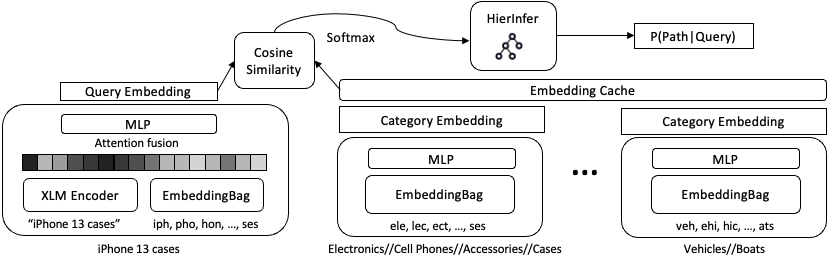}
\Description{Dual-encoder architecture of the HierCat model}
\caption{{\small Model architecture}}
\label{fig:modelArchitectureFig}
\vspace{-1.0em}
\end{figure*}

\subsection{Problem formulation}\label{sec:problem_formulation}
Facebook Marketplace uses a product taxonomy system called Facebook Product Taxonomy (FPT) to categorize product items. The taxonomy consists of around six thousand categories organized in a tree structure, with maximum depth equals to six.  The problem of query categorization is thus mapping a search query to the best category path on the taxonomy tree. Since a search query can be vague, a category path can end anywhere on the taxonomy tree and not necessarily on a leaf node. Formally, we model the conditional probability of a category path given a query $P(path | query)$, where $path$ is a sequence of category nodes on a taxonomy tree indexed by its level of the form $\{cate\_L_k, cate\_L_{k-1}, ..., cate\_L_1\}$, and $cate\_L_k$ is a category node at level $k$ of the taxonomy tree. To avoid modeling combinatorial number of taxonomy paths, we leverage the structure of the taxonomy tree and only model the conditional probability of a node given its ancestors using the following factorization
\begin{equation}
\begin{aligned}
& P(\{cate\_L_k,..., cate\_L_1\} \ |\ query) = \\
& P(cate\_L_k \ |\ \{cate\_L_{k-1},...,cate\_L_1\},  query) \ \cdot \\ 
& P(\{cate\_L_{k-1},...,cate\_L_1\} \ | \ query)
\end{aligned}
\end{equation}

In the following sections, we will use $path$ and $category$ interchangeably, both referring to the final category prediction like "Home//Furniture//Sofa". Sometimes we will use $node$ to refer to a node on the taxonomy tree (e.g. "Sofa").

\subsection{Weakly supervised training}\label{sec:data_collection}
We adopt a weakly supervised approach based on logged search engagement to train our classifiers. Specifically, we sample <query, product> pairs from 14 days of Facebook Marketplace's search engagement log on public content. All of the data are de-identified and aggregated, and are translated into 24 million <query, category> pairs based on a product item's category. <query, category> pairs with very low aggregation frequency are filtered out to reduce noise. Different predictions for the same search query are permitted because we hope to benefit from the richness of unfiltered information, especially since those labels may share some parent nodes (e.g. ,"Home//Furniture//Chair" v.s.  "Home//Furniture//Sofa"). Note that at Facebook Marketplace, a product item's category label can come from the seller, model prediction, or both. Labels from seller selection have limited coverage and are not very granular, but are accurate. Model predicted labels have better coverage but limited precision, especially at deeper levels. Note that this poses the challenge of heterogeneous training data for query categorization models.

\subsection{Pre-trained dual-encoder classification}
Instead of predicting query category as conditional probabilities as in section \ref{sec:problem_formulation}, we first treat categorization as a classic query classification problem for simplicity. In other words, we train a neural classification model $\Phi$ that predicts a node on the taxonomy tree given a query without considering the taxonomy structure. Formally, for a query and category node pair $(q, c)$, and the space of categories $C$, we minimize the cross-entropy loss defined as

\begin{equation}
  L(q, c_i) = -\log\frac{\exp(\Phi(q, c_i))}{\sum_{c_j \in C} \exp(\Phi (q, c_j))}
\end{equation}

For better generalization to different languages, we use a 2-layer XLM-encoder, a transformer-based language model pre-trained on large multilingual data \cite{xlm} to process query text. We also represent query text using character trigrams and encode it using an EmbeddingBag encoder \cite{pytorch}, as we find that multi-granular text representation with simple trigrams helps with short head queries. The two text representations are merged via a simple attention fusion layer as illustrated in figure \ref{fig:modelArchitectureFig}.

An interesting property of the query categorization problem is that the input query often shares the same text as its desired label. For example, query "electric boat" and category "Vehicle//Boat" both contain the word "boat". Motivated by this observation, we borrow the two-tower architecture commonly used in retrieval models \cite{que2search} to allow for an extra text encoder on the category labels and replace the final predication layer with the cosine similarities between query and category embeddings of dimension 128. Given that category text always come from a fixed set of six thousand categories, we only use EmbeddingBag encoder for the category tower to prevent over-fitting. 

\subsection{Product retrieval pre-training}
For the query embedding tower, we introduce an optional pre-training task based on embedding-based retrieval (EBR) of Marketplace products given search queries, because empirically, we discover that EBR model training generates nicely clustered search queries \cite{que2search}. For this work we adopt the Que2Search model \cite{que2search} which shares a very similar text encoder architecture and is trained to minimize the cosine similarities between engaged query and product pairs. We retrain the EBR model to ensure that there is no architectural difference.
    
\subsection{Hierarchical inference}
So far we treat query categorization as a flat classification problem. We discover this approach to be problematic due to the noise in user engagement data. For example, both "Cell Phones//Accessories//Cases" and "Cell Phones" are popular labels mined for query "iPhone 12", because a user could be browsing iPhone cases while shopping for an iPhone. Intuitively, both labels are useful to infer the query being a "Electronics" and "Cell Phones", but in the eyes of a normal multi-class classifier, they are just inconsistent labels.

To address this problem, we introduce a hierarchical inference algorithm to re-normalize the probability mass over a taxonomy tree. As illustrated in algorithm \ref{alg:hier_infer}, for all of the nodes on the taxonomy tree, we assign the cosine similarity generated from the dual-encoder model to $p\_cate[node]$. We then apply softmax over all of the leaf nodes to ensure they sum up to one. For any of the nodes at  leaf-1 level, we propagate the probabilities of its children back to itself, and then apply softmax at the leaf-1 layer. We repeat this iteratively at each layer until all of the levels are up-propagated and normalized. This ensures that the joint probability of all the nodes on a category path satisfies the conditional probability factorization we introduced in section \ref{sec:problem_formulation}. One caveat is that $p\_cate[node]$ still contains its original cosine similarity before probabilities from its children are added, and we believe that line 6 in algorithm \ref{alg:hier_infer} can be further generalized to be $\alpha * p\_cate[n] + SUM(p\_cate[CHILDREN(p\_cate[n])$ to control the impact of hierarchical inference vs. the original prediction. For simplicity of evaluation, we leave $\alpha = 1$ and did not tune the value in this paper.
\begin{algorithm}\label{hierInfer}
  \SetKwFunction{FMain}{hier\_infer}
  \SetKwProg{Fn}{Function}{:}{}
    \Fn{\FMain{$nodes$, $p\_cate$}}{
      Let $p\_cate[node] = cosine\_similarity(query, node)$ \\
      Let $levels = L[MAX\_LEVELS -1, \ldots, 0]$ \\
      \For{$l \in levels$}{
        \For {$n \in nodes[l]$} {
          $p\_cate[n] \mathrel{+}= SUM(p\_cate[CHILDREN(p\_cate[n])])$ \\ 
        }
        $p\_cate[nodes[l]] = SOFTMAX(p\_cate[[nodes[l]])$
      }
    }
\caption{Hierarchical inference}
\label{alg:hier_infer}
\end{algorithm}

%% file: experiments.tex
\begin{table*}[t]
\centering
\begin{tabular}{c|c|c|c|c|c|c}
    \toprule
    \small Technique & L1 F1 & L1 acc@5 & L3 F1 & L3 acc@5 & L6 F1 & L6 acc@5 \\
    \midrule
    \small FastText multi-label & 0.582 & 0.866 & 0.176 & 0.389 & 0.165 & 0.357 \\
    \small XLM classification & 0.762 & 0.869 & 0.339 & 0.416 & 0.233 & 0.315 \\
    \small XLM multi-label & 0.661 & 0.874 & 0.241 & 0.376 & 0.195 & 0.303 \\
    \small XLM classification + hierInfer & 0.765 & 0.907 & 0.344 & 0.426 & 0.233 & 0.315 \\
    \small Dual-encoder  XLM + hierInfer & 0.767 & 0.907 & \textbf{0.388} & \textbf{0.516} & \textbf{0.259} & \textbf{0.369} \\
    \small Dual-encoder  XLM +  hierInfer + trigram & \textbf{0.774} & \textbf{0.913} & 0.363 & 0.514 & 0.237 & 0.366\\ 
    \bottomrule
\end{tabular}
\vspace{2mm}
\caption{{\small Results for baseline comparisons and ablation studies}}
\label{tab:model_evaluation}
\vspace{-2.0em}
\end{table*}

\subsection{Experimental setup}
\textbf{Baseline models and ablation studies.} We perform ablation studies to evaluate each of our modeling choices and compare against alternative query classification methods commonly used across the industry. We compare our results against FastText multi-label classifier \cite{fasttext1, fasttext2} that assigns one class per-level as an hierarchical classifier. FastText is commonly used at Meta for short text classification and the multi-label approach also achieves good performance in recent query classification work \cite{aprfnet,backoffOptimize}. We also test directly fine-tuning an XLM model \cite{xlm}, a state-of-the-art pre-trained transformer model for classification. Since the XLM-classification model is used as part of our dual-encoder architecture, it also serves as an ablation study on the new architecture we propose. In addition, we experiment with combining the XLM-classification model with the multi-label approach as an alternative way to encode label hierarchy. 

\textbf{Data and evaluation metrics.} We train all of the models with the 24 million weakly supervised data described in section \ref{sec:data_collection}. For evaluation, we use a stratified sample of 173 thousand queries across different regions and languages (which were de-identified from any personal information of the users inputting the queries), and ask raters to map them to the right categories. We report micro-F1 and top five accuracy (true category matches with any one of the top five predictions) at different levels to access the model performances.

\textbf{Product retrieval pre-training.} We run separate experiments for adding product retrieval pre-training and report the relative improvement, because it is an optional step in our system that can work with any baseline architecture. For this ablation study we use the XLM + trigram query encoder.

\subsection{Ablation results}
As illustrated in table \ref{tab:model_evaluation}, dual-encoder architecture with hierarchical inference achieves the best performance across all levels, while hierarchical inference itself improves the performance on higher levels independent of the classification algorithm, because of its consolidation effect on higher categories by leveraging probabilities from their children. XLM classification model also outperforms FastText classifier possibly because of better generalization across languages \cite{xlm}. In addition, pre-training on product retrieval task, as shown in table \ref{tab:pretraining}, significantly improves leaf-level performance but does not help with L1. Our hypothesis is that leaf-level classification requires more granular embedding representations, and thus benefits from pre-trained embeddings more.

\begin{table}[tb]
\begin{tabular}{c|c|c|c|c|c}
    \toprule
    L1 F1 & L1 acc@5 & L3 F1 & L3 acc@5 & L6 F1 & L6 acc@5 \\
    \midrule
    neutral & neutral & +3\% & neutral & +16\% & +4\%   \\
    \bottomrule
\end{tabular}
\vspace{2mm}
\caption{{\small Improvements from product retrieval pre-training}}
\label{tab:pretraining}
\vspace{-3em}
\end{table}

%% file: online_testing.tex
\subsection{Serving-time optimizations}\label{system_arch}
We deployed {\systemname} on Meta's dedicated inference cloud \cite{predictor} and implemented several additional optimizations in the model's production path to improve latency and quality.

\textit{Embedding caching:} to speed up model inference, we only deployed the query encoder, and the category embeddings are pre-computed and cached. During inference, we first compute the query embedding, and then loop through the cached category embeddings to obtain the logits. Finally, hierarchical inference is performed to obtain the correct probability distribution over a taxonomy tree. A query level cache is also implemented to avoid duplicate inference calls for popular queries.

\textit{Beam search:} to further speed up inference and obtain a consistent taxonomy path, a beam search is performed by iteratively selecting the best children for each parent nodes. In production we simply use a beam size of one for simplicity. Beam search is terminated early if the score is lower than a threshold tuned from percentile method.

\begin{table}[tb]
\begin{tabular}{c|c|c}
    \toprule
    Technique & Engagement & NDCG \\
    \midrule
    L1 \& L2 category boost & neutral  & +1.4\% \\
    L3 category boost & +4.3\% & neutral \\
    \bottomrule
\end{tabular}
\vspace{2mm}
\caption{{\small Online A/B testing results for structured retrieval}}
\label{tab:structured_retrieval_results}
\vspace{-3.5em}
\end{table}

\subsection{Structured retrieval}
We leveraged {\systemname} in Facebook Marketplace Search to reduce the category mismatched results by emitting the per-level categories as optional retrieval terms, and boosting retrieval scores upon category matches. This is done with Meta's Unicorn system \cite{unicorn} but can also be achieved with open-source solutions like Elasticsearch \cite{elasticsearch}. The term weights are tuned by applying a linear regression to predict downstream ranking scores, and boosts are given to category matches up to level three. Two weeks of online A/B testing shows that such retrieval boost significantly improves both NDCG and online user engagement, as illustrated in table \ref{tab:structured_retrieval_results} (metric improvements in table \ref{tab:structured_retrieval_results} are all relative and incremental, and L3 category boost was launched after L1 \& L2 boost). Interestingly, we found that NDCG improved when L1 \& L2 boosts were added, while engagement stayed neutral until L3 boost was added. We discovered that this is due to discrepancies between the NDCG rating guideline and user behavior - while former focused more on top level category matches, it is the L3 match, which is the median of level distributions that is driving user engagement.

\subsection{Level distributions}
\begin{figure}[tb]
  \centering
  \includegraphics[width=0.4\textwidth]{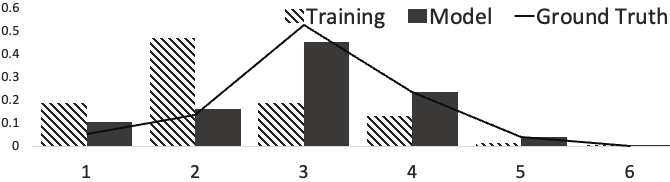}
  \Description{Level distributions of training data, model prediction and ground truth data}
  \caption{{\small Level distribution comparisons}}
  \label{fig:levelDist}
  \vspace{-2.0em}
\end{figure}

To understand the level of noise in the weakly supervised data mined from search engagement and how well our method is able to counter such noise, we examine the level distribution of query categories from logged online predictions, ground truth labels and training data as shown in figure \ref{fig:levelDist}. We can indeed observe a better overlay with the ground truth distribution from our production model, indicating that our method is able to de-bias its training data to be closer to ground truth.

%% file: related_work.tex



Query classification aims to classify search queries into a set of topical space to better understand users' intents.  This is generally treated as a text classification problem where the input texts are short and the class space is large. For example, Liu et al. \cite{Liu2019SystemDO} proposed an extreme classification approach using a mixture of CNN and Naive Bayes classifier based on such characteristics. In e-commerce, the label space is often a product taxonomy that possesses a tree structure, yet regular multi-class classification approach would either treat each node as a class, or only consider the the leaf categories \cite{aprfnet, dynmicQueryIntentJD, Liu2019SystemDO, Skinner2019EcommerceQC}. Liu et al. \cite{backoffOptimize} explored using multi-label FastText as a hierarchical classifier, and a back-off strategy to optimize the granularity-precision trade-off. Techniques to incorporate different dimensions of signals like contextual information \cite{dynmicQueryIntentJD, contextAwareQC} and search engine feedback \cite{aprfnet} are also explored to disambiguate search intent and augment training data.  In terms of the classification algorithm, popular text classification methods like XML-CNN \cite{xmlcnn} or FastText \cite{fasttext1, fasttext2} are often used as a component in a query classification system \cite{backoffOptimize, Liu2019SystemDO}. Weakly supervised labels mined from search logs are also explored as a data augmentation technique \cite{aprfnet, Liu2019SystemDO,dynmicQueryIntentJD}, while the characteristics of such data especially its implication to hierarchical classification is rarely discussed to the best of our knowledge.

%% file: conclusion.tex
In this paper we demonstrate {\systemname}, a query categorization system that learns from weakly supervised data by leveraging transformer encoders pre-trained on downstream product retrieval tasks, and the hierarchical structure of an e-commerce product catalog. We show the effectiveness of each modeling choice we made through ablation studies and comparing against popular baseline methods. To the best of our knowledge, {\systemname} is also the first to demonstrate the effectiveness of transformer-based dual-encoder architecture in e-commerce query categorization despite that it is a rather popular architecture in information retrieval tasks \cite{que2search}. We deploy {\systemname} on Facebook Marketplace Search, share our inference-time optimizations, and show through online A/B testing that it significantly improves NDCG and searcher engagement.